# Study of turning takeoff maneuver in free-flying dragonflies: effect of dynamic coupling

Samane Zeyghami[1], Haibo Dong[2]

Mechanical and Aerospace Engineering, University of Virginia, VA, 22902

## SUMMARY

**Turning takeoff flight of several dragonflies was recorded, using a high speed photogrammetry system. In this flight, dragonfly takes off while changing the flight direction at the same time. Center of mass was elevated about 1-2 body lengths. Five of these maneuvers were selected for 3D body surface reconstruction and the body orientation measurement. In all of these flights, 3D trajectories and transient rotations of the dragonfly's body were observed. In oppose to conventional *banked turn* model, which ignores interactions between the rotational motions, in this study we investigated the strength of the dynamic coupling by dividing pitch, roll and yaw angular accelerations into two contributions; one from aerodynamic torque and one from dynamic coupling effect. The latter term is referred to as Dynamic Coupling Acceleration (DCA). The DCA term can be measured directly from instantaneous rotational velocities of the insect. We found a strong correlation between pitch and yaw velocities at the end of each wingbeat and the time integral of the corresponding DCA term. Generation of pitch, roll and yaw torques requires different aerodynamic mechanisms and is limited due to the other requirements of the flight. Our results suggest that employing DCA term gives the insect capability to perform a variety of maneuvers without fine adjustments in the aerodynamic torque.**

## INTRODUCTION

Aerial maneuvers of different insect species are studied widely in the last decades and many aspects of maneuverability in insect flight are explored. Although insects are capable of performing a variety of aerial maneuvers, technical challenges in acquiring accurate three dimensional data limited diversity and depth of our knowledge about flapping flight in nature. Recently high speed photogrammetry technology together with different auto-tracking and reconstruction techniques (Wagner, 1986; Ristroph et al., 2009; Koehler et al., 2012) inspired more comprehensive and accurate studies on the different aspects of the insect flight. Aerial maneuvers of different insects have been recorded and detailed body and wing kinematics were measured (Wang et al., 2003; Ribak and Swallow, 2007). Maneuvering flight of fruit fly, for instance, has been extensively studied and many aspects of aerodynamics and dynamics of these flights are discovered (Fry et al., 2003; Bergou et al., 2010; Cheng et al., 2010). In free flight, predation and evasive flights have been studied, as well, which revealed significant information about population dynamics and individual fitness (McLachlan et al., 2003; Combes et al., 2012). But yet most of the studies on the dynamics of the insects' aerial maneuvers are either limited to tethered flight or narrowed down to the turns that are modeled mainly by conventional "banked turn" model. Similar to the fixed wing aircrafts, in banked turn, insect redirects flight force vector by banking to the direction of turn. Studies revealed that many insects and birds use this strategy for performing aerial turns such as fruit fly (Fry et al., 2003), locust (Berger and Kutsch, 2003), dragonfly (Alexander, 1986), fruit bats and cockatoos (Hedrick and Biewener, 2007; Iriarte-Díaz and Swartz, 2008). Banked turn is a semi-2D model mapped to the three dimensional nature of these maneuvers, derived from decoupling governing equations of motion (EOM). However some studies recorded aerial maneuvers of insects in which rotational velocities are simultaneously high (Schilstra and Hateren, 1999). Dynamics of these maneuvers cannot be modeled without considering dynamic coupling effects. But then it's not clear what role these couplings play in the maneuver and why insects employ both kinds of strategies.

---

[1] PhD student, sz3ah@virginia.edu
[2] Associate professor, Haibo.dong@virginia.edu

Since flight is an expensive type of locomotion, insects (and birds) have learned to use their wings, body and legs together to increase efficiency of their flight and enhance stability and maneuverability (Dudley, 2002). Although wings play the most important role in the flight, as the main source for aerodynamic forces and torque generation, insects know how to use their abdominal deflections to enhance flight stability (Dyhr et al., 2013). Several studies have shown that leg extensions play an important role in stabilizing the flight as well (Götz et al., 1979; Lorez, 1995). However, in our study of dragonfly's turning takeoff however we have shown another mechanism that dragonflies use to enhance maneuverability, together with using their four wings and abdominal deflections (abdominal deflections in these maneuvers are investigated separately in another study). Since dragonfly is taking off and changing the flight direction at the same time, extra mechanisms to execute the maneuver successfully and efficiently are possibly required.

In this manuscript we start with modifying governing equations of rigid body rotation consistent with wing-body configuration of dragonfly. Individual terms are then discussed and named. One of these terms, called DCA (dynamic acceleration coupling) which represents strength of the coupling in dynamics. More discussion on the physical interpretation of this term in the case of dragonfly's maneuver is provided later in this work (in discussion section). Afterward, a simple dynamic model is represented and used to examine effectiveness of the DCA term in manipulating the orientation. Argument is followed by representing free maneuvering flight measurements and calculating contribution of DCA term. Correlation of pitch and yaw velocity with this acceleration term is been examined. We finally discussed that dragonfly's body reorientation maneuver is rather a three dimensional phenomena than planar and decoupling EOM may result in inaccurate models. Later on in our discussion, we compared two maneuvers of the same dragonfly and showed how difference in the acceleration throughout the course of maneuvers can be explained by including DCA term in the analysis. We concluded that in spite of the more complex appearance of trajectories of 3D maneuvers compared to planar, dragonflies are able to perform a variety of 3D maneuvers employing a simpler control system by controlling the DCA term.

## MATERIALS AND METHODS

### Dragonflies

Dragonflies were captured during the summer months of 2010 and 2011 in Dayton, OH, and were identified as Eastern Pondhawk, (Erythimus Simplicicolli**s**) species. Before performing the experiments they were stored outdoor in separate containers which were open to sunlight and air. Table 1 includes morphological data of each individual. We will refer to the four individual dragonflies as blue (B), green (G), red (R) and yellow (Y). Dragonfly B is a male and all other three are females.

Table 1. Morphological data base of dragonflies in experiment

| Dragonfly | year | gender | Weight (mg) | Flapping frequency (Hz) | # of studied maneuvers | Total # of wingbeats |
|---|---|---|---|---|---|---|
| **B** | 2010 | Male | 265 | 41.7 | 1 | 3 |
| **G** | 2011 | Female | 83 | 26.3 | 2 | 7 |
| **R** | 2011 | Female | 166 | 33.3 | 1 | 5 |
| **Y** | 2011 | Female | 119 | 26.3 | 1 | 4 |

### Experimental setup and 3D reconstruction

The photogrammetry setup used for dragonfly image collection consists of three synchronized Photron FASTCAM SA3 60K high-speed cameras with 1024×1024 pixel resolution. They were aligned orthogonal to each other on an optical table and operated at 1000 Hz with at least a 1/20000 sec. shutter speed to capture the dragonfly flight videos. The dragonflies were illuminated by 3 halogen photo optic lamps (OSRAM, 54428). The cameras were positioned 1.5 meters away from the insects, giving a depth of field of 3-4 body lengths in all directions depending on the size of the specimen (Koehler et al., 2012). We arranged the light in shooting area in the left side of the insect and sit the insect on a paper stool (or on a rod in one case) in the middle of the shooting area. Asymmetric light stimulated a turning takeoff in all the insects of experiment. This method allowed us to shoot voluntary maneuvering flight of the dragonfly without tethering or interrupting the insect. Exceptionally, dragonfly B was shot one year earlier that we had a different light arrangement in the lab. As a comparison, we included this data in our analysis which is a free takeoff from a rod.

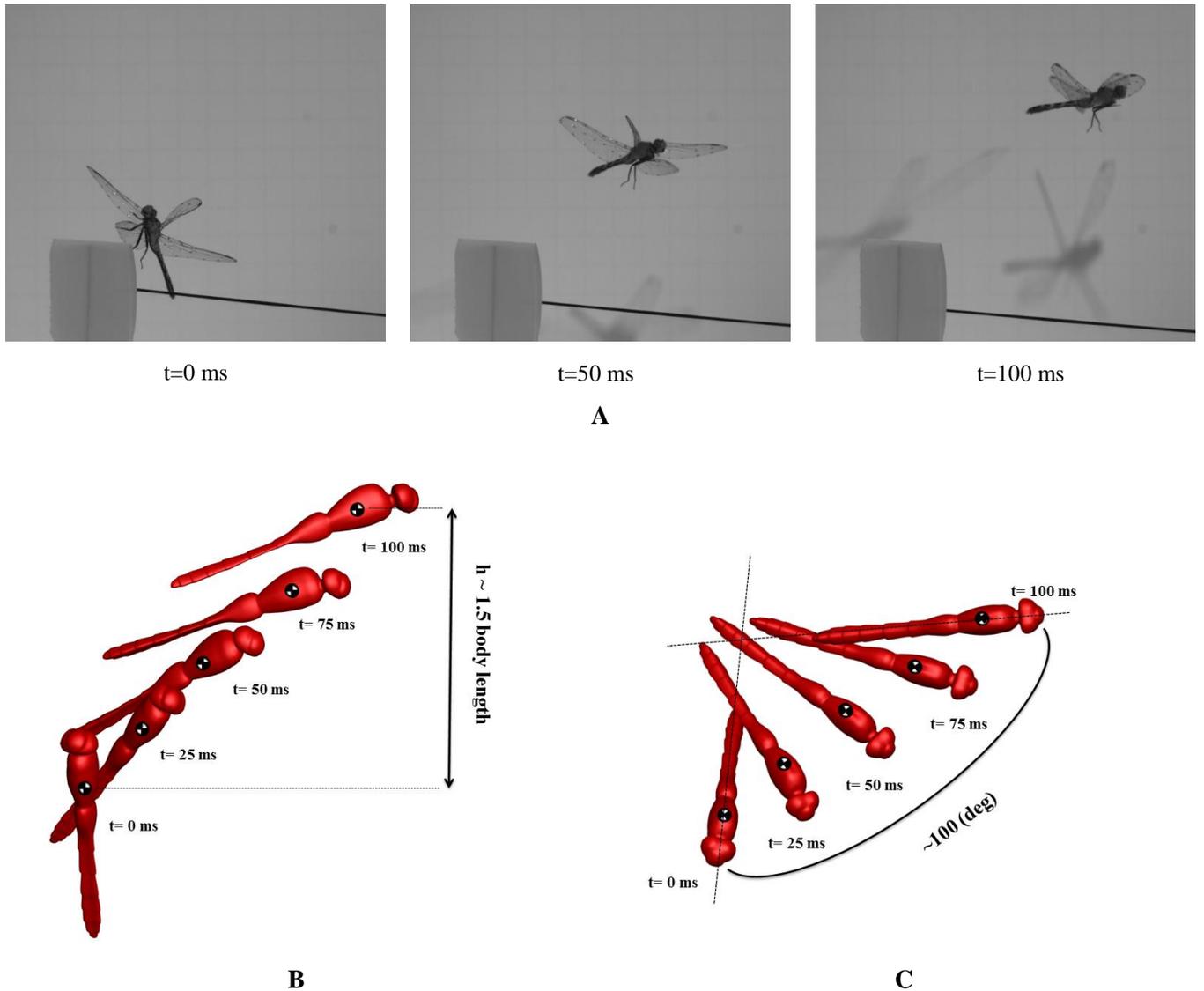

Fig. 1. A) Sample of 3 snapshots of a typical turning takeoff maneuver. (B) and (C) show reconstructed body of the same dragonfly in Fig. 1A. In a typical turning takeoff maneuver center of mass elevates about 1 to 2 body lengths and turn angle is about 80~120 degrees. Corresponding time is shown next to each snapshot.

The 3D surface reconstruction technique (Koehler et al., 2012) was applied to the raw output from the high speed cameras. We prepared a dragonfly-body template mesh which was animated by aligning the projection of its outer border to the silhouette of the body in the images. Body orientation and location in each video is measured by matching the wing roots' plane of the body mesh to the insect body every *1/1000* sec. consistent with the speed of the camera. We filmed each dragonfly several times and among those we chose videos in which at least two projections offered good quality images. Orientation of the wing roots plane in 3D space defines body orientation during maneuver. Normal vector to this plane is used to define the roll angle.

The internal consistency of the data further corroborates that the measurements were performed with adequate precision. Furthermore, our measured values were in good agreement with previously reported data on angular velocities of dragonflies in turn (Wang et al., 2003). Detailed descriptions about the reconstruction method and accuracy can be found in (Koehler et al., 2012).

### Description of turning takeoff maneuver

This terminology is used to describe the maneuver in which dragonfly redirects the flight while it's taking off at the same time. In addition to turn a significant elevation in center of mass, in order of a few body lengths, were observed during performance of this maneuver. Fig. 1 demonstrates a typical turning takeoff. Three sample frames taken directly from the high speed camera output (Fig. 1A). Fig. 1B-C show the reconstructed body at 5 snapshots of the same maneuver (as in Fig. 1A) in side and top views.

### Frames of reference and coordinate systems

Body orientation with respect to earth-fixed coordinate system is defined with three angles, bank ($\varphi$), elevation ($\theta$) and heading angle ($\psi$). Angular velocities, on the other hand, are usually studied about body axes and represented by roll velocity (P), pitch velocity (Q) and yaw velocity (R). Rotational velocities in body and earth-fixed coordinate systems are related as follows:

$$\dot{\varphi} = P + Q\sin(\varphi)\tan(\theta) + R\cos(\varphi)\tan(\theta)$$
$$\dot{\theta} = Q\cos(\varphi) - R\sin(\varphi) \qquad (1)$$
$$\dot{\psi} = (Q\sin(\varphi) + R\cos(\varphi))\sec(\theta)$$

Body coordinate system is right handed and fixed at the body center of mass which is estimated to be in the middle of the four wing roots on the longitudinal body axis. This axis points to the nose and is referred to x-axis. The y-axis points toward the right wings and the z-axis points downward.

### Equations of motion and effect of coupling term

Rotational motion of any rigid flying body can be expressed as follows;

$$\frac{d\vec{L}}{dt} + \vec{\omega} \times \vec{L} = \vec{M} \qquad (2)$$
$$\vec{L} = [I]\vec{\omega}$$

[*I*] is tensor of moment of inertia. On the right hand side of Eqn 2 is the vector of flight torque which in the case of insect's free flight is the same as aerodynamic torque. Inertial effects of the flapping wing also

have minor contribution in creating torque on the body but because of the very small wing to body mass ratio in dragonflies its effect is usually negligible (Hedrick et al., 2009). Inertial forces via fluid mass which is carried by wings are included in aerodynamic torque term. The first term on the left hand side is rate of generation of angular momentum (by the means of external torque applied on the system). The second term on the left hand is the cross product of instantaneous rotational velocity vector and angular momentum. In two dimensions both these vectors are normal to the plane of motion and thus their cross product is zero. Although further discussion on the physical meaning and importance of this term is postponed to discussion section, we will briefly show how it will appear in the definition of total acceleration. After expanding and rearranging terms in Eqn 2 and including effect of mass distribution of the dragonfly's body and wings, total rotational acceleration can be expressed as presented in Eqn 3. Detailed discussion and derivation are included in appendix A.

$$\dot{\omega} = \dot{\omega}_{aero} + DCA$$
$$\dot{\omega}_{aero} = [\frac{L}{I_{xx}} \quad \frac{M}{I_{yy}} \quad \frac{N}{I_{zz}}] \quad (3)$$
$$DCA = \left[ -\frac{\dot{R}+PQ}{2} \quad PR \quad -PQ \right]$$

We will refer to the two terms in right hand side of Eqn 3 as aerodynamic acceleration, $\dot{\omega}_{aero}$, and dynamic coupling acceleration (DCA) terms, respectively. DCA term is just a function of rotational velocities and accelerations and therefore can be directly measured in experiment. Relative magnitude of this term compared to aerodynamic acceleration reflects importance the dynamic coupling. Aerodynamic acceleration can be therefore calculated by subtracting instantaneous DCA vector from the total rotational acceleration.

**Dynamic model**

A conceptual dynamic model based on the Eqn A4 has been built to quantify potential effect of the DCA term in manipulation of the motion. Magnitudes of configuration constants (refer to appendix A) are calculated by measuring real insect weight and dimensions. Moment of inertia is calculated modeling head, thorax and tail of the insect as three ellipsoids while wings are modeled as two point masses located 40% of the wing length apart from the wing's root. We chose 40% rather than 50% (more natural choice) because of the denser venation structure which happens closer to the wing root. We then solved for pitch and yaw accelerations and consequently velocities as a function of pitch and yaw torques (inserted by fluid reaction to the wing's motion) and roll velocity. To investigate importance of the DCA term we assume that pitch and yaw torques are predefined and only DCA term can be controlled to alter the course of the maneuver. Although within wingbeat oscillations in flight torque may affect progress of the maneuver, for simplicity and without losing the generality of the model or conclusions, we assumed a constant pitch and yaw torque ($\tau_{pitch}$, $\tau_{yaw}$). On the other hand because DCA term is direct functions of roll velocity, a fairly more complicated model has been used for roll velocity, represented as summation of "m" harmonic functions.

$$P(t) = P_o + \sum_{n=1}^{m} P_n \sin(\frac{\Omega t}{n} + \alpha_{P_n}) \qquad (4)$$

$\Omega$ is the flapping frequency in Eqn 4. There are "3+2m" parameters which should be fixed to describe time course of yaw and pitch torques as well as roll velocity (Table 2). Magnitude of aerodynamic torques is chosen to be consistent with the measured values from dragonfly's flight. Keeping yaw and pitch torques unchanged we then calculated body orientation in several simulated maneuvers by changing parameters of roll velocity model in Eqn 4 (Fig. 2). Modulation in roll velocity (independent from yaw and pitch) manipulates the maneuver via altering the DCA term. We simulated four test cases by changing roll velocity amplitude. Reference maneuver is simulated without any roll velocity. In this case, aerodynamic torques result in about $100^o$ change in heading angle during a left turn (solid lines in Fig. 2).

Table 2. Aerodynamic torque and body morphology parameters in the dynamic model

| Parameter | $\tau_{pitch}(\mu N.m)$ | $\tau_{yaw}(\mu N.m)$ | Wingbeats | $I_{xx}(g.mm^2)$ | $I_{yy}(g.mm^2)$ | $I_{zz}(g.mm^2)$ | $I_{xz}(g.mm^2)$ |
|---|---|---|---|---|---|---|---|
| Value | 2 | -2.5 | 4 | 2.09 | 24.3 | 24.6 | -1.02 |

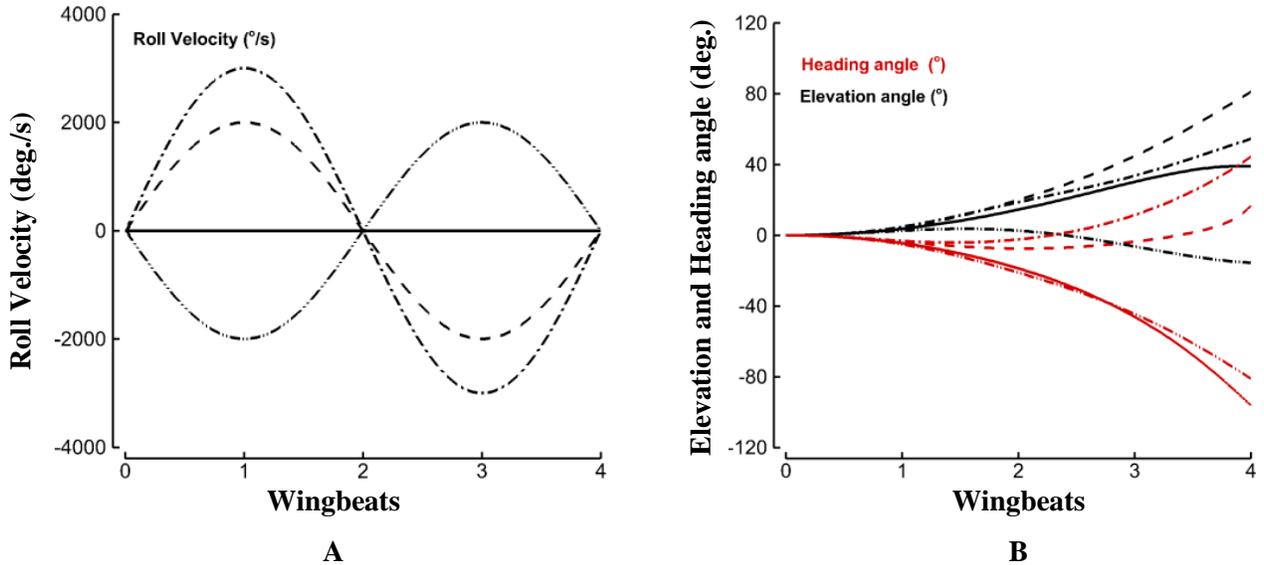

Fig. 2. A) Roll velocity is altered for each simulation by setting $m=4$ and varying amplitude in Eqn 4. It turned out that higher frequency terms in roll velocity doesn't affect the course of the maneuver significantly. Therefore we set $P_i$ {i=0,1,2,3} to zero. $P_4$ is respectively 0, 2000, -2000 and 3000 $^o$/s for solid, dashed, dashed double dot and dashed dot lines in (A) as well as in (B). The time course of change in the orientation is shown in (B). Red and black lines represent heading and elevation angles respectively.

In the next two test cases we increased roll amplitude to 2000 ($^o$/s) and then 3000 ($^o$/s) compatible with measured roll velocity in insect's maneuvers (Wang et al., 2003). Last test case has amplitude of -3000 ($^o$/s) which means the model insect will roll to the left first and then roll back to the right. All four cases resulted in totally different final orientations although yaw and pitch torques were identical for all of them. For instance in the third simulation, the model dragonfly ended up in a right turn although yaw torque was forcing it to turn left. Note that pitch angle is being affected as much as yaw. In the following sections we'll discuss how DCA term altered motion in real dragonflies' maneuvers.

## RESULTS
### Turning takeoff maneuver and patterns of change in body orientation

Filming takeoff of dragonflies with high speed camera revealed that in small time scales of few tenths of a second the insect performs a fast and controlled body reorientation maneuver. All the videos are reconstructed starting from the first frame that insect's fore legs detached from the stool or rod. Body orientation and rotational velocities are presented in Fig. 3. Turning takeoff maneuver was performed in 2 to 5 wingbeats in all cases. Heading angle changed between 30 to 120 degrees. All of the maneuvers initiated with a pitching up motion. In one of the cases which body was initially pitched up to about $80^o$, further pitch up motion of the body was negligibly small. $60-70^o$ elevation angle was preferred by dragonfly to perform the body reorientation maneuver (higher than typical $50^o$ preferred by most of the insect (Hedrick et al., 2009) in aerial turns).

**Dragonfly B.** This dragonfly was initially almost parallel to the horizon and banked $90^o$. Taking off from this situation can be hard due to orientation of the wings with respect to the gravity. Dragonfly B was able to reorient the flight in less than four wingbeats. Maneuver started with a fast roll to the left and followed by a pitch up, yaw to the right and rolling back to the right (Fig. 3, first row, solid lines).

**Dragonfly G.** We reconstructed two videos of dragonfly G. First one started from a comparatively lower initial elevation angle (about $30^o$). Maneuver started with a pitching up motion and a roll to the left. It followed by a roll to the right. Body was constantly yawing to the left during the maneuver. In second maneuver initial elevation angle was about $60^0$. Maneuver started with a pitch up and a fast roll to the left and yaw to the right. It immediately followed by pitching down and rolling to the right in second wingbeat. Body continued to pitch down and finally insect ended up with a $30^o$ elevation angle and more than a $100^o$ change in heading angle (Fig. 3, second and third row, solid lines).

**Dragonfly R.** Longitudinal body axis of this dragonfly was initially almost normal to the ground. This maneuver starts with a yaw motion to the left. Exceptionally in this flight, there is almost no pitch up motion at the beginning. Flight continues with a roll to the right and body pitches down until it ends up with about $40^0$ final elevation angle (Fig. 3, fourth row, solid lines).

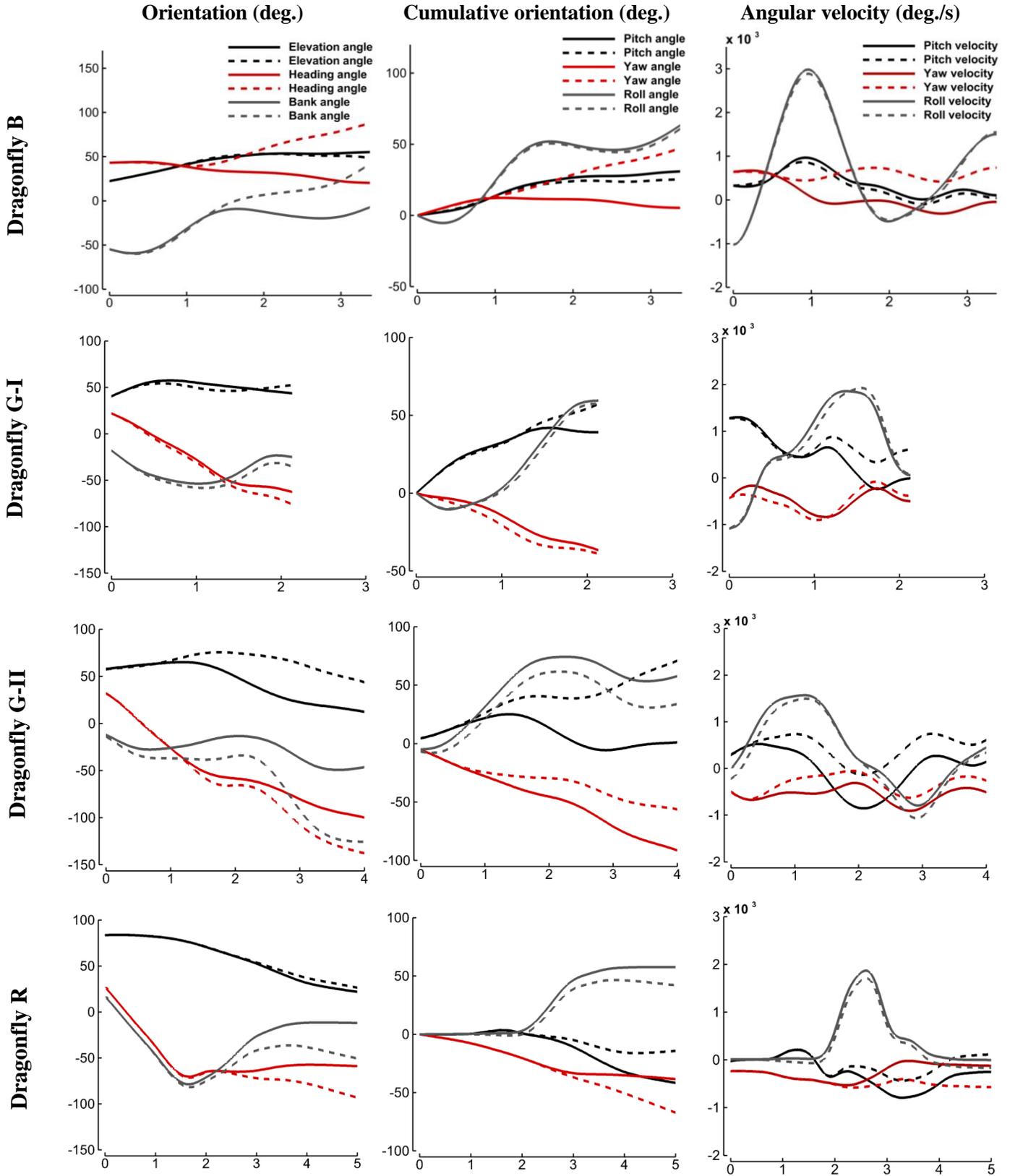

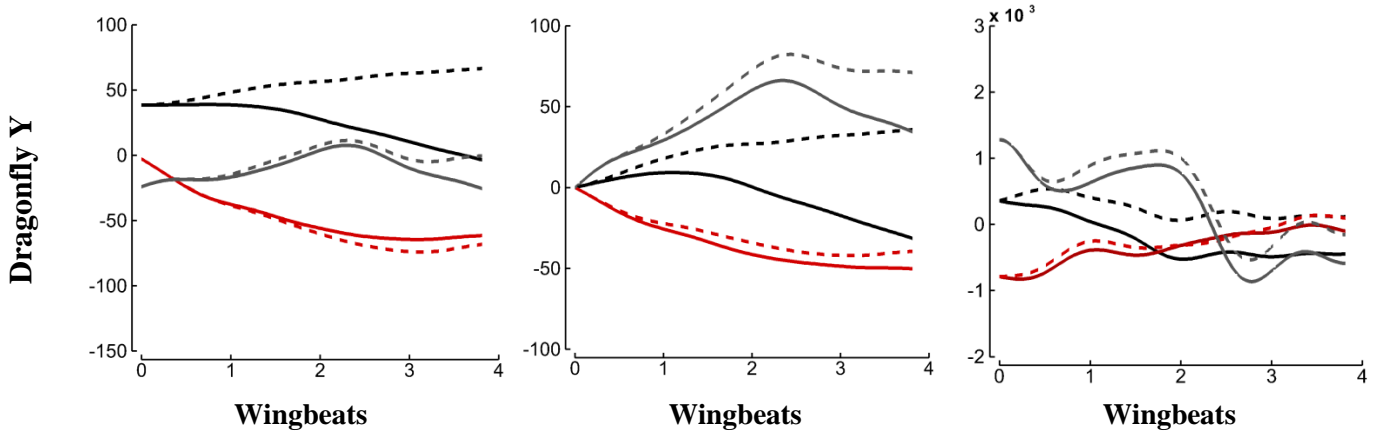

Fig. 3. Body orientation, cumulative body angular orienantaion and roational velocities about body axes are shown in first, second and third columns, respectively. All graphs are plotted versus wingbeat number and graphs on one column share the same legend. Solid lines show real insect maneuver measurments and dashed lines shows how motion will be if the coupling effect is neglected. Each row is tagged with the respective dragonfly's color as represented in Table 1.

**Dragonfly Y.** Initial elevation angle of this dragonfly was similar to the first maneuver of dragonfly G. Flight started with a slight pitch up motion and followed by a pitch down starting from second wingbeat. Body rolled to the right for more than three wingbeats and rolled back to the left afterwards. Constant yaw to the left was observed during the maneuver (Fig. 3, fifth row, solid lines).

Similar to other insects such as blowfly (Schilstra and Hateren, 1999) and compatible to other records of dragonfly's maneuvering flight (Wang et al., 2003) highest rotational velocities observed in roll. Maximum rolling velocity was about 2000 °/s in all flights except in B in which roll velocity were exceeded 3000°/s at the first wingbeat. This value decreased to about 1500$^0$/s at the second wingbeat. Pitch and yaw velocities both toped at about 1000°/s in some of the flights (Fig. 3, solid lines).

### Body orientating via dynamic coupling effect

To investigate how the maneuver is being affected by the DCA term, we calculated change in angular orientation without the coupling term by integrating aerodynamic acceleration term twice in time. Ignoring coupling term is equivalent to assuming that there is no significant interaction between the rotations (refer to appendix A for further explanation on how to separate aerodynamic acceleration from dynamic coupling by measuring rotational velocities and accelerations). We plotted this motion in the same plot with the measured body orientation (solid lines) in Fig. 3. Dashed lines show how the orientation will be if there was no coupling effect. Note that difference between the dashed and solid lines shows the effect of the DCA term in modulating the motion. We also plotted accumulated angular motion in body coordinate system which is calculated by integrating angular velocities about body axes. Coupling effect on changing body orientation was significant in all the maneuvers we recorded.

**Effect on angular velocities during the course of maneuver**

All the maneuvers start with a pitch up motion which lasts for about a wingbeat. It followed by a pitch down motion starting in second wingbeat during the maneuver. This behavior was the characteristic of all the turning takeoff maneuvers in this experiment. Although dragonflies are able to control flight torque independently about each of the three body axes (Taylor, 2001) adjusting all three of them simultaneously during the limited time course of the maneuver is a complicated task. Our analysis in previous section

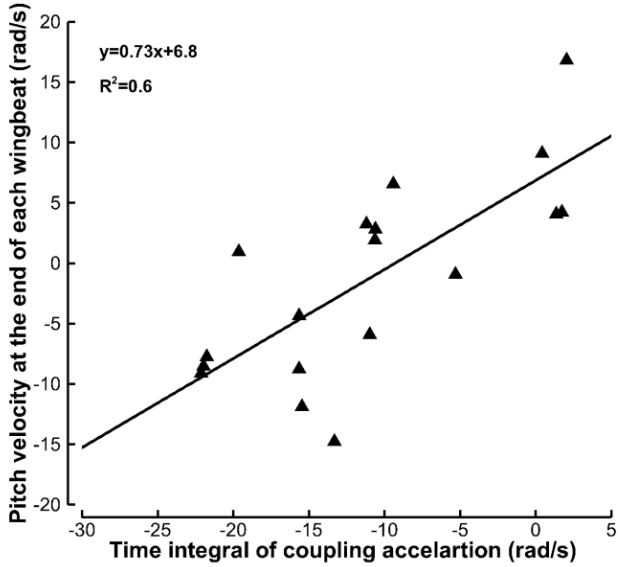

A-I

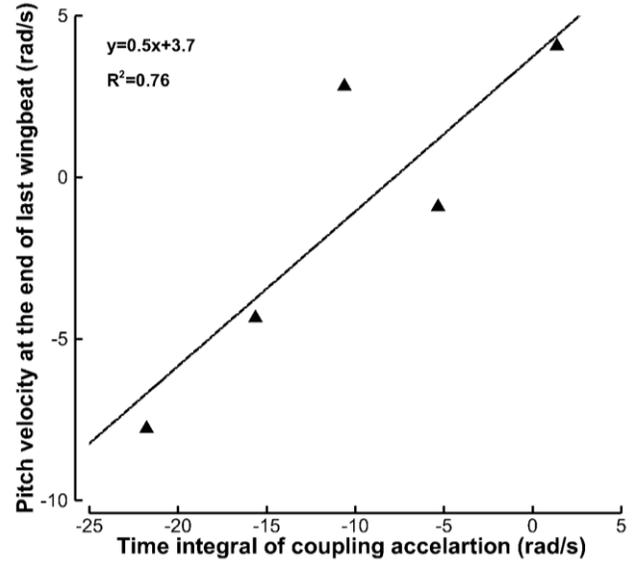

A-II

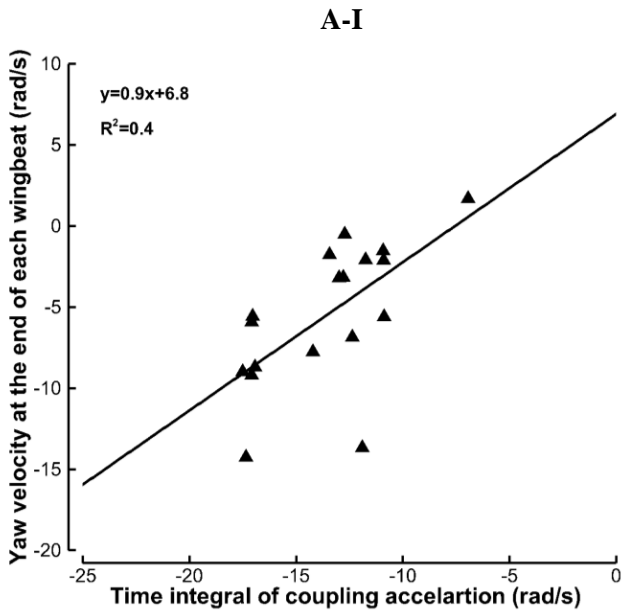

B-I

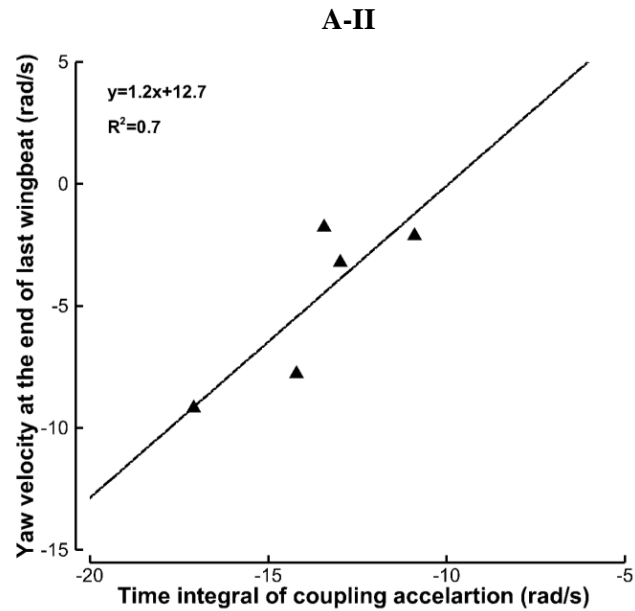

B-II

Fig. 4. Correlation of pitch (A-I) and yaw (B-I) velocities at the end of each wing stroke with time integral of DCA term. Correlation of the final pitch (A-II) and yaw (B-II) velocities with time integral of DCA term over the whole duration of the maneuver.

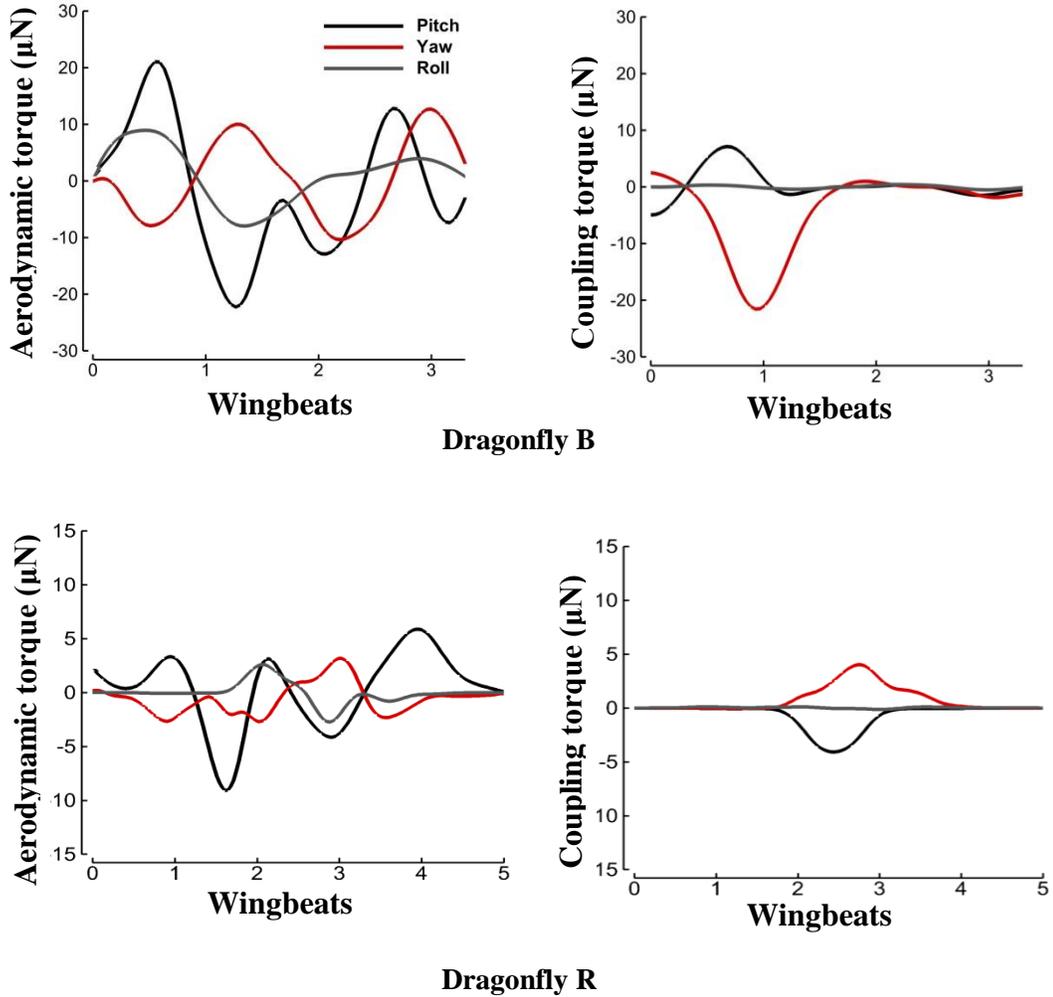

Fig. 5. Aerodnamic and coupling torques for two dragonflies B and R. All graphs share the same legend as shown in the top left of the figure.

shows that dragonflies G, R and Y always pitched body down with the help of DCA (Fig. 3). Similar effect was true about the yaw motion in dragonfly B which suggests existence of an interaction between yaw and pitch which can be controlled based on the maneuver requirements. Inspired by the observed behavior of the DCA term in stabilizing pitch, we examined correlation between the pitch velocity and DCA term using simple linear regression technique. A strong correlation was found between pitch velocity at the end of each wingbeat and time integral of DCA term in governing equation of pitch from beginning of the maneuver to the end of the respective wing stroke, $R^2=0.6$ (Fig. 4A-I). This correlation is even stronger with the final pitch velocity of each maneuver, $R^2=0.76$ (Fig. 4A-II).

Yaw was affected by DCA term significantly, as well. DCA mainly acted like a break for yaw motion as well as pitch except in dragonfly R's maneuver which yaw was actually extended by the help of the DCA. Similar to pitch, we found a strong correlation between yaw velocity at the end of each wingbeat and time integral of DCA from the beginning of the maneuver to the end of the respective wing stroke, $R^2=0.4$ (Fig. 4B-I). This correlation was even stronger with the final yaw velocity, $R^2=0.7$ (Fig. 4B-II).

**Equivalent coupling torque term**

A torque vector can be defined as the product of the DCA term and the moments of inertia. Importance of this term becomes clearer when we note that the difference between the three rotational motions is physical. The insect may use different mechanisms to roll, pitch or yaw the body. There are different limitations and capabilities involved with modifying each of these motions. Therefore, coupling torque term serves as an extra mechanism which offers an opportunity to accelerate/decelerate motion without using aerodynamic mechanisms that generate direct aerodynamic torques in the respective direct ion. For instance, yaw acceleration can be generated directly from the yaw torque term (N in Eqn A1) or can be carried out via coupling term. This effect can be significant in fast turn maneuvers of dragonflies. We plotted coupling torque in yaw and pitch in Fig. 5. Direct aerodynamic torque is also plotted in the same figure for comparison. Relative magnitude of the coupling with respect to the direct aerodynamic torque depends on the maneuver but our observation shows that it can even be the dominant instantaneous effect. Effect of the coupling torque is relatively bigger at the mid stage of maneuver in which a transition from acceleration to deceleration is expected to happen. Although there is no enough evidence that this term is intentionally controlled by the insect during maneuver, our results suggest that its magnitude is large enough to effectively modulate total flight torque.

**DISCUSSION**

**Extra mechanism to enhance pitch and yaw**

In oppose to planar (2D) motion, in three dimensional space, angular momentum vector is not aligned with angular velocity except in very rare geometries or occasions. Therefore, system can change configuration by tilting angular velocity vector (moment free rotation). But more importantly via modulating DCA, magnitude of this misalignment can be controlled and help modifying motion during the maneuver (Eqn 5).

$$\frac{d\vec{L}}{dt} = \vec{M} + \vec{L} \times \vec{\omega}$$

$$\vec{L} = \begin{bmatrix} I_{xx}P - I_{xy}Q - I_{xz}R \\ I_{yy}Q - I_{xy}P - I_{yz}R \\ I_{zz}R - I_{xz}P - I_{yz}Q \end{bmatrix} \quad (5)$$

Angular momentum vector is represented in its general form in Eqn 5, instead of the specified form for dragonfly's rigid body model, to emphasize that body bending may also contribute similarly to the maneuver by modulating angular momentum vector (via changing moment of inertia) and consequently the DCA term and angular acceleration. Therefore in addition to the aerodynamic torque, cross product of the instantaneous angular momentum and velocity can enhance the pitch or yaw, especially in the case of slender geometries like dragonfly's.

## Comparison between two maneuvers of dragonfly G

In two successive turning takeoff maneuvers, dragonfly G changed its heading about 90 and 120 degrees respectively (maneuver-I and II). If the effect of the coupling wasn't significant in these maneuvers, dynamics of the turn was expressible as bellow;

$$I_{xx}\dot{P} = L$$
$$I_{yy}\dot{Q} = M \qquad (6)$$
$$I_{zz}\dot{R} = N$$

In this case, rotational acceleration is directly derived from the aerodynamics torque. We compared mean rotational acceleration and net angular displacements in Table 3. In second maneuver dragonfly G yawed less than 40° while in the second one yaw exceeded 80⁰. The above model predicts that mean yaw torque in first maneuver is more than 4 times bigger than the second one (based on mean angular accelerations and Eqn 6). Similarly, mean pitch torque in first maneuver is predicted to be more than 16 times bigger than the second one. This means that for performing maneuver-I insect has to employ significantly higher levels of flight torques compared to maneuver II. On the contrary, our study suggests that main difference between these two maneuvers is generated via difference in the DCA term. We measured the mean DCA magnitude for each of the maneuvers and calculated mean aerodynamic acceleration based on Eqn 4. Results are summarized in Table 4. Our model suggests significantly lower ratios of aerodynamic torque between the two maneuvers (3.5 in pitch and 2.6 in yaw). Noting that duration of maneuver-I is about half of the maneuver-II, these ratios are reasonable. Consistent with these results Fig. 3 also shows that body was forced by flight torque to pitch up about $60^o$ and yaw about $50^o$ in both maneuvers. Thus, although two maneuvers are very different in the sense of the trajectory and the time course of the body orientation, aerodynamic torques in yaw and pitch are very similar in both maneuvers. Further modulation in motion is possible via modulation in the DCA term.

Table 3. Estimation of the mean pitch and yaw aerodynamic torques with and without DCA term for two maneuvers of dragonfly G, referred to as Maneuver-I and II. Last two columns show the ratio of pitch and yaw torques of maneuver-I and II.

|  |  | $\bar{\dot{Q}}(rad/s^2)$ | $\bar{\dot{R}}(rad/s^2)$ | $\bar{\dot{Q}}_{aero}(rad/s^2)$ | $\bar{\dot{R}}_{aero}(rad/s^2)$ | $\dfrac{\bar{M}_I}{\bar{M}_{II}}$ | $\dfrac{\bar{N}_I}{\bar{N}_{II}}$ |
|---|---|---|---|---|---|---|---|
| **Without DCA** | **Maneuver-I** | -273 | -14 | - | - | **17.0** | **3.5** |
|  | **Maneuver-II** | -16 | -4 |  |  |  |  |
| **With DCA** | **Maneuver-I** | -131 | -55 | -141 | -70 | **3.5** | **2.6** |
|  | **Maneuver-II** | 23 | 30 | -40 | 26 |  |  |

### Dynamic coupling effect offers simpler control during maneuvers

Strong correlation between rotational velocities and DCA term implies that the effect of this term should be considered in study of performance of a turning takeoff maneuver. Simultaneous rotational motions may occur due to aerodynamic coupling effects. For instance, increasing thrust on one wing to perform a yaw increases lift as well, and roll the body simultaneously. Although current research does not consider the aerodynamic mechanisms and their respective coupling effects, we measured the motion accurately and calculated how much torque was needed for such maneuvers to be performed. We were able to measure DCA as well since it's just a function of rotational velocities (Eqn 3). Our results show that three rotational motions of body have more dynamic interactions than what is currently assumed. Further investigation showed that the magnitude DCA is strongly correlated with rolling velocity ($R^2$=0.8). We interpreted that as work done by pitching torques can accelerate/decelerate yaw if the body rolls fast enough. Our model along with measurements suggest that body rotation can be even reversed without modulation in yaw torque using existing pitch velocity of the system and manipulating roll velocity as desired. Since body response in roll is much faster than yaw and pitch, this method almost immediately modulates acceleration of these two rotations. Since accurate manipulation about each of the body axes is an expensive control task, this method offers significant degree of control over the motion during the maneuver while costing less control effort. It worth to be remarked here again that our results suggest that dragonflies utilize DCA to accelerate rotation about the specific body axis as well as employing it as an effective break to settle pitch and yaw motions down. We also showed that magnitude of the corresponding torque to coupling acceleration is significant enough to be used as a flight torque modulation tool.

Clearly, rotational velocities have significant effect on the aerodynamic torque generation as well. For instance, recorded roll velocities are high enough to create significant angle of attack difference between the wings on the two sides of the turn and that can contribute in accelerating or decelerating the maneuver.

This effect, which can be significant for dragonflies because of their high aspect ratio wings, will be studied and quantified in the future study.

**Flapping frequency and its connection with maneuvering behavior**

To perform a maneuver two important time scales contribute to the performance. The first one is related to the required time to modulate flight forces and torque, and the second is correlated with the body's dynamic response time. In flapping flight, period of flapping, T, characterizes required time for modulating force and torque. Moreover, it is reasonable to describe response time of the body by the timing of the maneuver, $t_{mnvr}$. When flapping frequency is high (low T/ $t_{mnvr}$ ratios) insect is able to modulate flight force and torque instantaneously compared to the timing of the maneuver. It is expected that in this case, the contribution of the phenomena which occurs within a wingbeat can be abstracted into the resulting stroke-averaged torque. Thus, the dynamics of the maneuver can be explained by the average flight torque and body inertial properties ( $I\ddot{\alpha} = N$ ). But although there is no significant variation in the duration of the maneuver of flying insects (they are in the range of hundreds of milliseconds) flapping frequency ranges widely from a few to a few hundred Hz. As the ratio of the flapping frequency to the maneuver time (T/ $t_{mnvr}$) increases, within wingbeat dynamics may couple with the overall dynamic of the turn. In this case contribution of within wingbeat phenomena expands to time history of the adjustments in flight torque (compared to stroke-averaged in the case of high flapping frequency flight) and within wingbeat changes in body motion. Therefore, simultaneous rotation about all body axes maybe even unavoidable. In this study we showed that to explain dynamics of the dragonflies' turning takeoff maneuvers, DCA term should be considered in the analysis. Further than that our results suggest that DCA is employed by insect to control the maneuver mainly as a break to both pitch and yaw motions. Further investigation is needed to prove whether similar effects are observable in other flapping fliers with high T/t ratios or not.

**List of symbols and abbreviations**

| | |
|---|---|
| $\phi$ | Bank angle |
| $\theta$ | Elevation angle |
| $\psi$ | Heading angle |
| $p$ | Roll velocity |
| $q$ | Pitch velocity |
| $r$ | Yaw velocity |
| $I$ | Tensor of moment of inertia |
| $\vec{L}$ | Angular momentum vector |
| $\dot{\omega}_{aero}$ | Aerodynamic acceleration term |
| $DCA$ | Dynamic coupling acceleration term |

| $\Omega$ | Flapping period |
| $\tau$ | Flight torque |
| $R^2$ | Coefficient of determination in simple linear regression model |

## Appendix A

Governing equations of rigid body rotation, Euler equations, can be expanded as in Eqn A1. On the left hand side of this equation there are configuration parameters such as mass and moment of inertia, along with inertial forces and torques.

$$\begin{aligned} I_{xx}\dot{P} - I_{xz}\dot{R} - I_{xz}PQ + (I_{zz} - I_{yy})RQ &= L \\ I_{yy}\dot{Q} + (I_{xx} - I_{zz})PR + I_{xz}(P^2 - R^2) &= M \\ I_{zz}\dot{R} - I_{xz}\dot{P} + (I_{yy} - I_{xx})PQ + I_{xz}QR &= N \end{aligned} \quad (A1)$$

P, Q and R are rolling, pitching and yawing velocities and dot notation shows the first derivative with respect to the time. These terms represent angular velocities about body axes. L, M and N are rolling, pitching and yawing torque and are referred to as "direct aerodynamic torque" in this paper. To get a clearer picture, we rearranged Eqn A1 as below,

$$\begin{aligned} \dot{P} &= \frac{L}{I_{xx}} + C_P RQ + D_P (PQ + \dot{R}) \\ \dot{Q} &= \frac{M}{I_{yy}} + C_Q PR + D_Q (R^2 - P^2) \\ \dot{R} &= \frac{N}{I_{zz}} + C_R PQ + D_R (\dot{P} - QR) \end{aligned} \quad (A2)$$

For the sake of convenience we defined two matrices C and D which are defined based on the wing-body configuration as bellow;

$$C = \begin{bmatrix} C_P \\ C_Q \\ C_R \end{bmatrix} = \begin{bmatrix} \frac{I_{yy} - I_{zz}}{I_{xx}} \\ \frac{I_{zz} - I_{xx}}{I_{yy}} \\ \frac{I_{xx} - I_{yy}}{I_{zz}} \end{bmatrix}, \quad D = \begin{bmatrix} D_P \\ D_Q \\ D_R \end{bmatrix} = \begin{bmatrix} \frac{I_{xz}}{I_{xx}} \\ \frac{I_{xz}}{I_{yy}} \\ \frac{I_{xz}}{I_{zz}} \end{bmatrix} \quad (A3)$$

Moment of inertia of the insect is being estimated by modeling head, thorax and tail with three ellipsoids (details can be found in (Zeyghami and Dong, 2012)). Since body-wing configuration is symmetric about the longitudinal axis $I_{xy}=I_{yz}=0$. Note that if the body-wing is symmetric about lateral plane (plane which divides body to upper and lower sections) $I_{xz}$ becomes zero similarly. Usually in insects wings are very light compare to the body (in the case of dragonflies less than 1% of the body weight) and body shape is very close to cylindrical which results in smaller roll moment of inertia compared to the moment of inertia about yaw or pitch axes. $I_{xz}$ and $I_{xx}$ terms are negligible compared to $I_{yy}$ and $I_{zz}$. We accurately estimated body moment of inertia for dragonfly B (Table A1). To obtain moment of inertia of the other individuals based on the values obtained for dragonfly B, we scaled these values by the ratio of product of mass and body length squared, assuming that body shape is pretty similar for all dragonflies of the same species. Note that in calculating moment of inertia we assumed that wings are widely spread on the sides of the body in a plane that passes through the center of mas of the system. Our calculations, show that governing dynamics of the rotational motion of a dragonfly in the turning takeoff maneuver reduces to,

$$\dot{P} = \frac{L}{I_{xx}} + \frac{(\dot{R} + PQ)}{2}$$
$$\dot{Q} = \frac{M}{I_{yy}} + PR \qquad (A4)$$
$$\dot{R} = \frac{N}{I_{zz}} - PQ$$

Table A1. Weight, dimensions and MOI for wings and each of the body sections of dragonfly B.

|  | Weight (mg) | Roll MOI (mg.mm$^2$) | Yaw MOI (mg.mm$^2$) | Pitch MOI (mg.mm$^2$) |
| --- | --- | --- | --- | --- |
| **Head** | 1.8 | 60.3 | 1520.0 | 1540.0 |
| **Thorax** | 184.6 | 705.0 | 685.00 | 7020.0 |
| **Tail Part I** | 3.1 | 29.5 | 720.0 | 728.0 |
| **Tail Part II** | 2.1 | 15.1 | 15300.0 | 15300.0 |
| **Fore Wings** | 3.0 | 670.0 | 20.0 | 22.6 |
| **Hind Wings** | 3.0 | 610.0 | 5.3 | 5.3 |

**AKNOWLEGEMENT**

We would like to thank our colleagues Chris Koehler and Ning Zhe for their help in 3D reconstruction of the videos and Chengyu li for his help in preparing some of the pictures. This work is supported by NSF CBET-1055949.